\documentclass[10pt,journal,compsoc]{IEEEtran}
\ifCLASSOPTIONcompsoc
  \usepackage[nocompress]{cite}
\else
  \usepackage{cite}
\fi
\ifCLASSINFOpdf
\else
\fi
\usepackage{graphicx}
\usepackage{booktabs}
\usepackage{color}
\usepackage{url}
\usepackage{xspace}
\usepackage{adjustbox}
\usepackage{multirow}
\usepackage{array}
\usepackage{caption}
\usepackage{subcaption}
\usepackage{relsize}

\newif\ifcommentvar
\commentvartrue

\ifcommentvar
	\newcommand{\mei}[1]{\textcolor{black}{#1}}
	\newcommand{\reza}[1]{\textcolor{purple}{#1}}
	\newcommand{\gema}[1]{\textcolor{red}{#1}}
	
\else
	\newcommand{\mei}[1]{}
	\newcommand{\reza}[1]{}
	\newcommand{\gema}[1]{}
\fi

\begin{document}
%
\title{On the Relationship Between the Developer's Perceptible Race and Ethnicity and the Evaluation of Contributions in OSS}
%
%
%
%

\author{Reza~Nadri,
        Gema~Rodr\'iguez-P\'erez,~\IEEEmembership{Member,~IEEE,}
        Meiyappan~Nagappan,~\IEEEmembership{Member,~IEEE,}
\IEEEcompsocitemizethanks{\IEEEcompsocthanksitem All the authors are associated with David R. Cheriton School of Computer Science, University of Waterloo, Waterloo, ON, Canada.
\protect\\

E-mail: \{rnadri, g5rodrig, mei.nagappan\}@uwaterloo.ca}}

%
%

\markboth{IEEE TRANSACTIONS ON SOFTWARE ENGINEERING}%
{Shell \MakeLowercase{\textit{et al.}}: Bare Demo of IEEEtran.cls for Computer Society Journals}
%



\IEEEtitleabstractindextext{%
\begin{abstract}
\textbf{Context:} Open Source Software (OSS) projects are typically the result of collective efforts performed by developers with different backgrounds. Although the quality of developers' contributions should be the only factor influencing the evaluation of the contributions to OSS projects, recent studies have shown that diversity issues are correlated with the acceptance or rejection of developers' contributions. \textbf{Objective:} This paper assists this emerging state-of-the-art body on diversity research with the first empirical study that analyzes how developers' perceptible race and ethnicity relates to the evaluation of the contributions in OSS. We also want to create awareness of the racial and ethnic diversity in OSS projects. \textbf{Methodology:} We performed a large-scale quantitative study of OSS projects in GitHub. 
We extracted the developers' perceptible race and ethnicity from their names in GitHub using the Name-Prism tool and applied regression modeling of contributions (i.e, pull requests) data from GHTorrent and GitHub. \textbf{Results:} We observed that (1) among the developers whose perceptible race and ethnicity was captured by the tool, only 16.56\% were perceptible as Non-White developers; 
(2) contributions from perceptible White developers have about 6-10\% higher odds of being accepted when compared to contributions from perceptible Non-White developers; and (3) submitters with perceptible non-white races and ethnicities are more likely to get their pull requests accepted when the integrator is estimated to be from their same race and ethnicity rather than when the integrator is estimated to be White. \textbf{Conclusion:} Our initial analysis shows a low number of Non-White developers participating in OSS. Furthermore, the results from our regression analysis lead us to believe that there may exist differences between the evaluation of the contributions from different perceptible races and ethnicities. Thus, our findings reinforce the need for further studies on racial and ethnic diversity in software engineering to foster healthier OSS communities.
\end{abstract}


}

\maketitle
\section{Introduction}
\label{sec:intro}
In any line of work, diversity regarding race, gender, personality traits,
or age is beneficial beyond ethical reasons~\cite{avery2012there,galinsky2015maximizing,reynolds2017teams}. Particularly, in Software Engineering \emph{(SE)}, diversity helps to address a problem from different perspectives, designs more robust software products, and seems to create more efficient teamwork~\cite{earley2000creating}. Indeed, diversity has been recognized as a high-value team characteristic~\cite{vasilescu2015gender,pieterse2006software,catolino2019gender} and many companies have increased their efforts to create more diverse teams.\footnote{https://diversity.google/}\footnote{https://www.microsoft.com/en-us/diversity}\footnote{https://diversity.fb.com/read-report/}

More than ever, diversity can be seen in online collaborative coding platforms such as GitHub. GitHub attracts developers from all around the world to contribute to Open Source Software~\emph{(OSS)} development. Usually, the collaborative development cycle in GitHub starts with a developer submitting a contribution, e.g., a pull request. Then, a project developer with the right permissions evaluates the developer's pull request to accept or reject the contribution. Through this paper, we will refer to a developer submitting a pull request as the~\emph{Submitter} and a project developer evaluating the pull request as the~\emph{Integrator}. 

It is believed that the integrator evaluates the pull requests based on code quality or factors related to the quality of the source code being contributed~\cite{scacchi2007free}. However, pull request quality is not the only factor to consider when evaluating pull requests in GitHub. Recent studies have demonstrated that specific diversity factors related to social~\cite{tsay2014influence}, personality~\cite{iyer2019effects}, gender~\cite{terrell2017gender}, and geographical location~\cite{rastogi2018relationship} may affect the acceptance or rejection of pull requests. For example, Terrell~\emph{et al.}~\cite{terrell2017gender} found that among those who have identifiable gender outside the project, females have lower acceptance. Rastogi~\emph{et al.}~\cite{rastogi2018relationship} identified statistically significant differences in pull request acceptance rate (the number of accepted pull requests over the number of submitted pull requests) between developers from different countries. Recently, Iyer~\emph{et al.} showed that developers' personality traits correlate with pull request acceptance~\cite{iyer2019effects}. 

Moreover, Vasilescu~\emph{et al.}'s GitHub Survey~\cite{vasilescu2015perceptions} highlighted that developers are aware of the ethnicity of their team members; and that 30\% of GitHub developers have felt sometimes negative experiences due to diversity in terms of their national origin, language, and ideology. Although previous works have studied the relationship between different non-technical factors on pull request acceptance~\cite{tsay2014influence,vasilescu2015gender,terrell2017gender,rastogi2018relationship,iyer2019effects}, developer's race and ethnicity has not been examined yet. These previous findings along with the lack of studies about open source developers' race and ethnicity as a non-technical factor in pull request evaluation motivate this study. 

In this paper, we study the perceptible race and ethnicity in the OSS community
(1) to create awareness of the racial and ethnic diversity distribution of OSS developers in GitHub; and (2) to understand the differences between pull request acceptance of different perceptible racial and ethnic groups. This study is essential because if there are imbalanced distributions between developers' perceptible races and ethnicities groups; and there are differences in the acceptance rate of developers from different perceptible races and ethnicities, this study recognizes the problem and may create a responsibility to increase racial and ethnic diversity in OSS and GitHub communities. Moreover, understanding the relationship between developers' perceptible race and ethnicity and the evaluation of their pull request may indicate that OSS developers in GitHub experience the conflicts that have existed between racial and ethnic groups through history. 

To shed light on the racial and ethnic diversity in OSS, we carried out a large-scale empirical analysis in GitHub that answers the following research questions:

\textbf{RQ0:} \textit{ How many developers in each perceptible race and ethnicity are successfully participating in OSS using GitHub?}
\textbf{Motivation:} The proportion of people belonging to different races and ethnicities is not reflected in the OSS development process. This lack of racial and ethnic diversity in the development process can have undesired consequences as the software tools, processes, and final software products might not be inclusive. Therefore, answering this question helps creating awareness of the presence or lack of racial and ethnic diversity in the OSS community.

\textbf{RQ1:} \textit{To what extent is the submitter's perceptible race and ethnicity related to the acceptance probability of a pull request?}
\textbf{Motivation:} When an integrator can infer the race and ethnicity of the submitter from the submitter's name, the integrator may make judgments based on intuitions or internal stereotypes. Since the OSS community strives to be a meritocracy~\cite{fielding1999shared,schmidt2001leveraging}, it should avoid any unconscious or unintentional racial and ethnic bias when it exists. Thus, answering this question is essential to find any empirical evidence that can help the GitHub and OSS community to understand perceptible racial and ethnic factors that might limit the OSS community to behave as a true meritocracy.

\textbf{RQ2}: \textit{To what extent is the submitter-integrator perceptible race and ethnicity pair related to the acceptance probability of a pull request?}
\textbf{Motivation:} Integrators can infer the submitter's perceptible race and ethnicity and then identify themselves as being of the same race and ethnicity of the submitter. Note that we will use the term estimated race and ethnicity for the integrator since they know their race and ethnicity. However, we can only estimate the race and ethnicity of the integrator with the NamePrism tool. 
Based on this perception of belonging to the same group, integrators might be more likely to accept pull requests from submitters perceptible as their same race and ethnicity~\cite{byrne1971attraction,tajfel1982social}. If this happens, 
Based on existing research on similarity attraction~\cite{byrne1971attraction}, and social identity and social categorization~\cite{tajfel1982social}, OSS integrators in GitHub may have an unconscious or unintentional bias towards these submitters that should be avoided. This RQ allows us to test whether existing research of affinity bias predicts the outcomes of decisions in the Software Engineering/SE domain.

We first identified developers' race and ethnicity based on their names using the Name-Prism tool~\cite{ye2017nationality}, which has an F1 score of 0.795. When we refer to submitters' race and ethnicity, we prefix it with ``perceptible'' as  NamePrism estimates is what is likely perceived by others. When we refer to integrators' race and ethnicity, it is not longer perceptible as the integrator knows what their ethnicity is. Hence, we prefix integrators' race and ethnicity with the word ``estimated'' since this is the race and ethnicity estimated by the NamePrism tool. We then used GHTorrent alongside GitHub's Application Programming Interface to extract pull requests and related features to link them with their respective developers. We used regression techniques from past studies on the relationship between non-technical factors and pull request acceptance ~\cite{vasilescu2015gender,terrell2017gender,rastogi2018relationship} to build regression models to assess the effect of submitters' perceptible race and ethnicity on how likely it is for a pull request to be accepted (pull request acceptance probability). 

Through this longitudinal empirical study, we have analyzed more than two million pull requests from $37,762$ projects and $365,607$ developers in GitHub. Our findings show a relationship between submitters' perceptible race and ethnicity and pull request acceptance probability that can be used as a feature to predict whether a pull request gets accepted. They also reveal that developers who are perceptible as White have a higher acceptance rate (the number of accepted pull requests over the number of submitted pull requests), and they suggest that being perceptible as a White submitter increases 6-10\% the odds to get a pull request accepted when compared to being perceptible as an API or a Hispanic submitter. Finally, we found that pull requests from perceptible API, Hispanic, and Black submitters are more likely to get accepted when the integrators' estimated race and ethnicity is the same as the submitters' perceptible race and ethnicity (0.36, 0.75, and 9.5 times higher odds when compared to estimated White integrators, respectively).

The primary contributions of our paper include:
\begin{itemize}
    \item We empirically observed the relationship between the perceptible race and ethnicity of the submitters and the evaluation of their pull requests in OSS projects in GitHub.
    \item We demonstrate that there are differences in pull request acceptance rate among different perceptible racial and ethnic groups.
    \item We show that the submitter's perceptible race and ethnicity can affect pull requests acceptance probability, controlling for other variables.
    \item We identified a low number of perceptible Black, AIAN, and 2PRace submitters and integrators who contribute to OSS projects in GitHub.
\end{itemize}

The rest of the paper is organized as follows. Section~\ref{sec:rel_work} discusses the background and related work. Section~\ref{sec:meth} presents our case study design, including the data collection, our dependent variable, various independent variables, and the statistical modeling. Section~\ref{sec:res} answers our research questions and shows the findings of our study. Section~\ref{sec:disc} discusses the results. Section~\ref{sec:thre} highlights the threats to validity, and Section~\ref{sec:conc} concludes the paper and discusses future work.

\section{Related Work}
\label{sec:rel_work}
\subsection{Theoretical background}

Ethnicity is often related to cultural expressions (i.e., religion, beliefs, and customs) and identification~\cite{cornell2006ethnicity}. The race is related to physical characteristics such as skin color, which have been erroneously attributed to the DNA~\cite{duster2009debating,cosmides2003perceptions}. Although we acknowledge that ethnicity and race are different concepts, through this paper, we use the term ``race and ethnicity'' to refer to the categories Black or African American, Hispanic or Latinx, Native American (AIAN), Asian Pacific Islander (API), and White because we used the U.S. Census categorizations for identifying the perceptible race and ethnicity of developers.

Decades of social studies have demonstrated that race and ethnicity is an influencing factor for the career success of individuals. African-American people are likely to earn less when compared to White people~\cite{council1998changing}. Furthermore, when applying for jobs, perceptible African-American applicants need to send almost double the number of resumes than perceptible White applicants to get one callback~\cite{bertrand2004emily}. 

This discriminatory behavior can be explained by some social psychology theories which say that working in groups tends to trigger discriminatory behavior against individuals that are not a member of the group~\cite{byrne1971attraction,tajfel1982social}. For instance, \emph{Similarity-Attraction theory (SA)} postulates that people prefer working with others similar to them~\cite{byrne1971attraction} and \emph{Social Identity and social Categorization theory (SIC)} indicates that people tend to categorize themselves into groups~\cite{tajfel1982social}. These theories suggest that members of one's group are treated better than outsiders. Moreover, psychological research on dual-process theory claims that individuals use two different systems of thinking when making impressions and judgments~\cite{evans2003two}. One system is slower and more deliberate, while the other is based on an individual's intuition or gut-feeling. This second gut-feeling system often becomes involved when there is enough available information about the target that activates an individual's stereotypical expectations.

GitHub is a collaborative environment where different OSS developers from different racial and ethnic groups can work together to develop and maintain software. In GitHub, the integrators analyze pull requests and make decisions based on different technical and non-technical factors~\cite{rastogi2018relationship,terrell2017gender,tsay2014influence}. Factors such as the perceptible race and ethnicity derived from the submitters' name, the GitHub profile picture of submitter, or other additional information available can trigger integrators' unintended inclinations during pull request evaluation process.

Therefore, with this study, we want to understand whether there is any evidence suggesting that some racial and ethnic groups may prefer contributions from individuals of their same group rather than contributions from outsiders. Specifically, we study whether the perceptible race and ethnicity of a submitter can predict whether the pull request is accepted or not. Notice that our study does not analyze the exact reasons behind the existence of unconscious bias, social stereotypes, or other inclinations. But the results might serve as a starting point to further understand the unconscious bias and social stereotypes in OSS development.

\subsection{Study of diversity and social factors in Software Engineering}
To the extent of our knowledge, our paper is the first empirical study that identifies the diversity regarding perceptible race and ethnicity on OSS and analyzes the racial and ethnic demographics of OSS developers in GitHub; and also the relationship between developers perceptible race and ethnicity and pull request evaluation. Previously, Vasilescu \emph{et al.}~\cite{vasilescu2015perceptions} surveyed GitHub developers to identify their perceptions of diversity on GitHub. Their results indicated that 30\% of developers are aware of the ethnicity of their fellow developers and that this awareness is statistically significant between females and males. However, Vasilescu \emph{et al.}'s study~\cite{vasilescu2015perceptions} did not study whether there is a relationship between developers' perceptible race and ethnicity and pull request evaluation.

Recent studies have addressed other diversity issues in OSS contributions.
Calefato~\emph{et al.}~\cite{calefato2019large} studied the personality of developers and categorized developers into three personality types. They found that personality traits are not changing over time, or with changing roles. Vasilescu~\emph{et al.}~\cite{vasilescu2015gender} identified the gender imbalance in OSS and found that gender diversity has a positive correlation with team productivity. Terrell~\emph{et al.}~\cite{terrell2017gender} found when the gender is identifiable and the developers are from outside a project, men have a higher acceptance rate comparing to women. Catolino~\emph{et al.}~\cite{catolino2019gender} studied the gender of developers in OSS and the effect of gender imbalance on community smells.

In another body of work, researchers have tried to understand pull request acceptance process and the factors that may influence pull request evaluation.
Tsay~\emph{et al.}~\cite{tsay2014influence} showed that project managers use not only technical factors but also social clues while evaluating pull requests. Prior interactions inside the project and ``social distance'' were important to pull request acceptance process. Gousios~\emph{et al.}~\cite{gousios2014dataset} studied the factors affecting pull request acceptance on 1.9 million pull requests. Their results reaffirmed the existence of non-technical factors involved in pull request evaluation process. Iyer~\emph{et al.}~\cite{iyer2019effects} found that pull requests from developers who are more open and conscientious, but less extroverted, have a higher likelihood to be approved. They also found that developers who are more conscientious, extroverted, and neurotic have a higher likelihood of accepting a pull request. Rastogi~\emph{et al.}~\cite{rastogi2018relationship} added geographical location to previous studies and studied 17 countries that have at least 1\% of the total number of pull requests. They found that country of residence can influence pull request acceptance. They also found that when the submitter and the integrator are in the same country, the chance of pull requests getting accepted is higher. Our study is different from Rastogi~\emph{et al.}'s study because their study relies only on the current country of the developers and does not address any form of racial and ethnic diversity in the OSS community. Indeed, Rastogi~\emph{et al.}'s study cannot be seen as a race and ethnicity study because inferring races and ethnicities from countries is not an accurate approach as there might be different races and ethnicities coexisting in a country. We see this in our data as well, the majority of developers in Nigeria are perceptible as Black developers, however, there are also perceptible White developers in the country (see Figure~\ref{fig:heatmap}). Similarly, in the U.S. would be considered as perceptible White developers and would ignore all the perceptible Black, Hispanic, and API developers there. Thus, inferring race and ethnicity from countries would be biased towards the majority racial and ethnic group in a country.

\section{Methodology}
\label{sec:meth}
To measure the effect of perceptible race and ethnicity on pull request acceptance, we mined data from projects in GitHub. We used GHTorrent \cite{Gousi13} alongside GitHub's developers API to extract data. We collected data from projects, users, and pull requests. We then selected a subset of the dataset to continue our study. We finally used the collected data to build regression models for analysis. 


\subsection{Project Selection}
Although GitHub has $125,486,232$ projects and more than $52$ million pull requests\footnote{ According to GitHub's data publicly available on June 2019}, not all of the projects are interesting to study. To make sure that we excluded trivial projects (e.g. homework assignments) from our analysis, we only selected a subset of the projects. To obtain this subset, we used 
a publicly-accessible dataset\footnote{\url{https://reporeapers.github.io/results/1.html}} that contains the score of a repository based on best engineering practices. This dataset was obtained using the reaper tool~\cite{munaiah2017curating}, which assesses a GitHub repository in the form of a score. This tool uses score-based and random-forest classifiers (trained on two datasets, organization and utility dataset) to determine whether a project is non-trivial, which outperforms other approaches with high precision (82\%) and high recall (86\%). We chose projects with more than ten stars, which at least three classifiers had classified them as non-trivial because we believe that if the projects meet these criteria, they are likely to be engineered software projects. It is important to study engineered software projects because we want to study the population of software developers and not necessarily anyone else who may collaborate on the GitHub platform. 


\subsection{Pull Request Selection}
\label{sec:pr_selection}
We inferred a pull request acceptance using the pull request's status. The status is collected directly from the GitHub API. A pull request's status can be either open, merged, or not-merged (rejected). We used the merge time field to determine whether a pull request is merged (accepted). We considered a pull request as not-merged when the merge time is null and the pull request is closed. We considered a pull request as merged when the merge time is not null and the pull request is closed. Otherwise, when the pull request's status is not closed, we considered the pull request as open. Since cherry-picked pull requests do not have a specific merge time, these pull requests were considered as not-merged. In total we extracted $4,029,190$ pull requests.

We filtered out the pull requests where the same person is both the submitter and the integrator because these cases can introduce bias in our dataset. We also excluded open pull requests from our analysis because it may be accepted or rejected in the future. $1,521,599$ pull requests were removed by these filters. We labeled $2,039,601$ as merged, $467,990$ as not-merged.

Table~\ref{tab:raw_data} shows the number of projects, pull requests, and developers in GitHub\footnote{at June 2019}, after applying the projects' selection criteria and the pull request's selection criteria. After the second filter, our dataset of submitters, integrators and projects are representative of the population of GitHub submitters, integrators and projects that could have been chosen for this study and to which the results of this analysis should generalize. 

\begin{table}
\centering
  \caption{Number of projects, pull request, and developers identified in GitHub, after the first filtering (Section 3.1), and after the second filtering (Section 3.2).}
  \label{tab:raw_data}
  \begin{tabular}{llll}
    \toprule
    Number of & GitHub & 1st Filter & 2nd Filter\\
    \midrule
    Projects & $125,486,232$ & $46,191$& $37,762$\\
    Pull requests & $52,018,443$ & $4,029,190$ & $2,507,591$\\ 
    Developers & $32,411,734$ & $493,170$ & $365,607$\\ \hline
  \bottomrule
\end{tabular}
\end{table}

\subsection{Deriving race and ethnicity from names} 
\label{sec:eth_selection}
We relied on the registered name of developers in GitHub to estimate their perceptible race and ethnicity. In GitHub, the developer's name is an optional field. Therefore, developers can enter any valid characters as names. We started with $493,170$ developers. To maximize the accuracy of our models, we estimated developers' perceptible race and ethnicity using these tools:

\textbf{(1) Stanford Named Entity Recognizer \emph{(NER)}}: First, we used the Stanford NER~\cite{finkel2005incorporating} to discover whether a set of characters includes names. In general, Stanford NER is a model that takes a set of names and labels, each of them as a class such as a person, organization, protein, etc. Stanford NER is a classifier based on a linear-chain Conditional Random Field. There are multiple versions of Stanford NER for different classes and different languages. In this paper, we used English Stanford NER with three classes. We used this tool to classify developers' names as either person, organization, or location. Our dataset only includes developers with at least one name labeled as a person. This step recognized $320,633$ inputs as names.
    
\textbf{(2) Name-Prism}: Second, we used Name-Prism~\cite{ye2017nationality} to infer the perceptible race and ethnicity of developers, using their names. Name-Prism introduces name-embedding and utilizes the concept of homophily to create a name-based perceptible nationality/ethnicity classification tool. Name-embedding converts each name to a vector and tries to recognize contexts and similarities of names in the same context. The context in the case of name-embeddings is perceptible ethnicities/nationalities. Homophily is a term used in communication sciences, which alludes to the fact that people tend to communicate with similar people. Name-Prism uses this phenomenon in the context of instant messaging, i.e., people from a racial and ethnic group tend to communicate with other people from the same racial and ethnic group. By combining these two concepts and collecting 74M labeled names from 118 countries, Junting~\emph{et al.} created the most accurate classification tool to identify races and ethnicities with an F1 score of 0.795~\cite{ye2017nationality}. The second best classifier \textit{Ethnea} \cite{torvik2016ethnea} has only an F1 score of 0.580. Based on U.S. Census Bureau, Name-Prism uses six racial and ethnic groups: White, Black, Hispanic, API (Asian, Pacific Islander), AIAN (American Indian and Alaska Native), and 2PRACE (Mixed Race) to build the classifier. It produces a confidence rate between 0 and 1 for each group. Name-Prism could identify race and ethnicity of all names but with different confidence levels, $282,312$ with more than 0.8 confidence rate. \footnote{We have chosen a threshold of 0.8 based on a sensitivity analysis of the tool. This analysis is included as an online appendix (A) in our reproducibility package~\cite{repro}}. When Name-Prism could not predict a perceptible race and ethnicity with more than 0.8 of confidence, we classified this developer's perceptible race and ethnicity as ``Unknown''. The ``Unknown'' category also has developers without full names in their profiles and developers for which the NER tool failed.


\subsection{Deriving developers' geographical location} 
\label{sec:location_selection}
We also extracted the developers' geographical location following the approach proposed by Rastogi~\emph{et al.}~\cite{rastogi2018relationship} and the ``country-NameManager'' script provided by Vasilescu~\emph{et al.}~\cite{vasilescu2015gender}. 

\subsection{Feature Selection}
\label{sec:fe_selection}
To explain the relationship between perceptible race and ethnicity, and pull request acceptance, we first need to find out what features or characteristics are correlated with pull request acceptance. Prior work has grouped these features into three categories: project's characteristics, developer's characteristics, and pull request's characteristics~\cite{gousios2014dataset}. This categorization is based on prior work in the areas of bug triaging, developer recommendation, pull request, and patch acceptance studies. The features that we have extracted are chosen from three sources \cite{gousios2014dataset,tsay2014influence,rastogi2018relationship}. The project level features that need access to the source code (e.g. number of test cases or tests lines per logical lines of code) are not included in this study because we have concentrated mainly on developers' and pull requests' characteristics. We collected features from pull requests and their respective actors and projects. Table \ref{tab:pr_features} shows our collected features vs. features introduced in similar studies. To obtain the features, we have two primary sources, the GitHub's API and GHTorrent public dataset (until June 2019). For the features that are common with Rastogi~\emph{et al.} (including country) \cite{rastogi2018relationship}, we applied their methodology and we used their provided scripts to collect these features. 

To identify main and outsider members of a repository, we followed Rastogi~\emph{et al.}'s approach~\cite{rastogi2018relationship}. We identified main developers as developers who have merged or closed a pull request on that repository, and the rest of the developers are outsiders on that repository.

\begin{table*}
\caption{Independent Variables}
\label{tab:pr_features}
\renewcommand{\arraystretch}{1.3}
\centering
\begin{tabular}{lll}
\hline
    \textbf{Feature} & \textbf{Literature} & \textbf{Description}  \\\hline \toprule \\
    \multicolumn{3}{c}{\textbf{Project Characteristics}} \\\hline \bottomrule 
repo\_popularity     & \cite{tsay2014influence,gousios2014dataset,rastogi2018relationship} & \begin{tabular}[c]{@{}l@{}} Numerical variable that shows the popularity of the repository at the time of \\the pull request's submission. Measured using the number of stars. Extracted from GHTorrent. \end{tabular} \\ \hline
repo\_team\_size      &\cite{tsay2014influence,gousios2014dataset,rastogi2018relationship}& \begin{tabular}[c]{@{}l@{}} Numerical variable that indicates the number of users associated with the repository \\in any way. A proxy for measuring the repository's size. Extracted from GHTorrent. \end{tabular}   \\ \hline
repo\_maturity       & \cite{tsay2014influence,rastogi2018relationship} & \begin{tabular}[c]{@{}l@{}} Numerical variable that shows how long (in months) the repository has existed before \\the pull request in pull request evaluation. Extracted from GHTorrent. \end{tabular} \\\hline
repo\_external\_contribs    &\cite{gousios2014dataset,rastogi2018relationship}& \begin{tabular}[c]{@{}l@{}} Numerical variable that indicates the percentage of the contribution made by users outside \\the repository's community. Extracted from GHTorrent.\end{tabular}  \\ \toprule \\
\multicolumn{3}{c}{\textbf{Submitter Characteristics}} \\ \hline   \bottomrule
prs\_main\_member  &\cite{tsay2014influence,gousios2014dataset,rastogi2018relationship}& \begin{tabular}[c]{@{}l@{}} Binary variable that indicates whether the submitter is a main member of the repository.\\ Extracted from GHTorrent. \end{tabular}   \\ \hline
prs\_popularity               &\cite{tsay2014influence,gousios2014dataset,rastogi2018relationship}& \begin{tabular}[c]{@{}l@{}}Numerical variable that indicates the popularity of the submitter. Measured by the number \\of submitter's followers at the time of pull request's submission. Extracted from GHTorrent.\end{tabular}  \\ \hline
prs\_watched\_repo     &\cite{tsay2014influence,rastogi2018relationship}& \begin{tabular}[c]{@{}l@{}}Binary variable that indicates whether the submitter and the repository have had prior \\association. Whether the submitter watches the repository. Extracted from GHTorrent.\end{tabular}  \\ \hline
prs\_succ\_rate             &\cite{gousios2014dataset,rastogi2018relationship}& \begin{tabular}[c]{@{}l@{}} Numerical variable that indicates the past success of submitter when submitting a pull request \\in GitHub. It is measured by the number of accepted pull requests divided by the total number \\of pull requests made by the submitter, on GitHub.  Extracted from GHTorrent\end{tabular}  \\ \hline
prs\_tenure  &\cite{rastogi2018relationship}& \begin{tabular}[c]{@{}l@{}} Numerical variable that indicates how long the user has been registered on GitHub at the time \\of pull request's submission. Extracted from GHTorrent. \end{tabular}  \\ \hline
prs\_experience             &\cite{gousios2014dataset,rastogi2018relationship}& \begin{tabular}[c]{@{}l@{}} Numerical variable that indicates the GitHub pull requests experience of the submitter, measured \\using the number of pull requests made by the submitter on GitHub.  Extracted from GHTorrent\end{tabular}\\ \hline
prs\_repo\_experience &       & \begin{tabular}[c]{@{}l@{}} Numerical variable that indicates the experience of the submitter in a GitHub repository. Measured by \\the number of pull requests submitted by the same submitter in the same repository before this pull request.\\ It captures the GitHub pull request experience of the submitter in the repository gained through time. \end{tabular}\\ \hline
same\_nationality &\cite{rastogi2018relationship}& \begin{tabular}[c]{@{}l@{}} Categorical variable that indicates whether the submitter and the integrator reside in the ``same'' \\ country or in ``different'' country. If the country data is unavailable for the submitter or integrator,\\ this variable is ``Unknown''. \end{tabular} \\ \hline
prs\_continent & \cite{rastogi2018relationship} & \begin{tabular}[c]{@{}l@{}} This categorical variable indicates the Submitter's continent of residence.  It was identified \\using the country of residence found in the user's profile~\cite{rastogi2018relationship}. If the country of residence \\is not available, this variable is ``Unknown''. Extracted from GHTorrent. \end{tabular}  \\ 
\toprule \\
\multicolumn{3}{c}{\textbf{Pull Request Characteristics}} \\ \hline \bottomrule
pr\_files\_changed &\cite{tsay2014influence,gousios2014dataset,rastogi2018relationship}& \begin{tabular}[c]{@{}l@{}} Numerical variable that indicates the number of files changed by the pull request. A proxy \\to measure pull request's size.  Extracted from GitHub API.\end{tabular}  \\ \hline
pr\_num\_comments &\cite{tsay2014influence,gousios2014dataset,rastogi2018relationship}& \begin{tabular}[c]{@{}l@{}} Numerical variable that indicates the number of comments on the pull request. A proxy to \\measure the importance of the pull request.  Extracted from GitHub API.\end{tabular}  \\ \hline
pr\_intra\_branch    &\cite{gousios2014dataset,rastogi2018relationship}& \begin{tabular}[c]{@{}l@{}} Binary feature that indicates whether the pull request was made intra branch.\end{tabular}  \\ \hline
pr\_num\_commits     &    \cite{gousios2014dataset,rastogi2018relationship}       & \begin{tabular}[c]{@{}l@{}} Numerical variable that indicates the number of commits made by the pull request. A proxy to \\measure pull request's size.  Extracted from GitHub API. \end{tabular} \\ \toprule \\
\multicolumn{3}{c}{\textbf{Race and Ethnicity Characteristics }} \\ \hline \bottomrule
prs\_eth   &   & Categorical variable that indicates the submitter's perceptible race and ethnicity. \\ \hline
pri\_eth  &   & Categorical variable that indicates the integrator's perceptible race and ethnicity. \\ \hline
\end{tabular}
\end{table*}

\subsection{Statistical Modeling}
\label{sec:modeling_selection}

To analyze the relationship of a submitter's perceptible race and ethnicity to pull request acceptance probability, we have used the submitter's perceptible race and ethnicity ($prs\_eth$) as an independent variable. Since we could not infer the race and ethnicity for all submitters and integrators, and in order to compute the relationship of perceptible race and ethnicity on pull request acceptance reliably, we further excluded pull requests in which their submitter's perceptible race and ethnicity or integrator's estimated race and ethnicity is ``Unknown''. We also removed AIAN perceptible race and ethnicity from the dataset because there was only one data point.

To the previous independent variable, we have added some control variables: the submitter's repository experience ($prs\_repo\_experience$), the submitter's continent of residence ($prs\_continent$) and previous features from the projects, the pull requests, and the submitters, that have been previously identified as possibly influencing pull request acceptance~\cite{gousios2014dataset,rastogi2018relationship,tsay2014influence}. Note that we have identified 197 different countries in our dataset, some of them with only one or two data points (making them ``Rare Events''). This unbalanced variable might lead to an unstable regression model, insignificant coefficients, and skewed predicted probabilities \cite{KinZen01}. Thus, we decided to add the developers' continent of residence instead of the developers' country of residence. We discuss the trade off of this decision in Section~\ref{sec:disc}.


To answer RQ1, we combined the previous variables and built a mixed-effect logistic regression model. We selected mixed-effect models instead of logistic regression models because they can capture measurements from within the same group (i.e., within the same project and within the submitters' group) as random effects~\cite{bates2014lme4}. We are interested in comparing the effect of perceptible race and ethnicity on pull request acceptance, but we are not particularly interested in the effects of the specific submitters and projects (in our dataset) on pull request acceptance. Thus, we used the identity of submitters and projects as random effects. To build the mixed-effect model we used the generalized linear mixed-effects model~\cite{bates2014lme4} function (glmer) available in the R package lme.~\footnote{\url{https://cran.r-project.org/web/packages/lme4/lme4.pdf}}


To answer RQ2 we have set White as the default value for the integrator's estimated race and ethnicity and we have created three subsets from our dataset: (1) pull request submitted by perceptible Hispanic developers and closed by estimated White integrators and estimated Hispanic integrators; (2) pull request submitted by perceptible API developers and closed by estimated White integrators and estimated API integrators; and (3) pull request submitted by perceptible Black developers and closed by estimated White integrators and estimated Black integrators. Then, we added a new variable $same\_ethnicity$ to the data in each subset. This variable is a categorical variable that is 1 when the integrator's estimated race and ethnicity is the same as the submitter's perceptible race and ethnicity, and 0 when it is not the same. For example, in the subset with perceptible Hispanic submitters, $same\_ethnicity$ is 1 when the integrator is also estimated to be Hispanic and 0 when the integrator is estimated to be White. Finally, for each subset, we built a mixed-effect logistic regression model and analyzed the effect of the $same\_ethnicity$ variable.

Note that we have not included the pull request closed by estimated API and Black integrators in the first subset because only 0.1\% and ~0\% of the pull requests were submitted by perceptible Hispanic developers and closed by estimated API and Black integrators, respectively. Similarly, we have not included the pull request closed by estimated Hispanic and Black integrators in the second subset of the data and the pull request closed by estimated Hispanic and API integrators in the third subset of the data.

After building the models for RQ1 and RQ2, we analyzed the Variance Inflation Factors \emph{(VIF)} to detect multicollinearity among independent variables. \emph{VIF} is calculated for each variable, and it is ranged from 1 upwards, lower value means lower multicollinearity. Previous studies~\cite{qiu2019going} in SE have chosen to remove variables with $VIF$ values higher than 5 or higher than 2~\cite{casalnuovo2015developer}. We chose an intermediate $VIF$ value of 3. Finally, coefficients were considered important if they were statistically significant ($\rho < 0.05$). The overall magnitude of the effect for our race and ethnicity variable was obtained using the ANOVA statistical test~\cite{cuevas2004anova}.


\section{Results}
\label{sec:res}
\subsection{RQ0: How many developers in each perceptible race and ethnicity are successfully participating in OSS using GitHub?}
\label{sec:rq0}

\begin{table*}
\centering
\renewcommand{\arraystretch}{1.2}
\caption{Races and Ethnicities population description}
\label{tab:eth_dist}
\begin{tabular}{l|cl|cl|cl}
\toprule
         & \multicolumn{2}{c|}{Number of Developers} & \multicolumn{2}{c|}{Number of pull requests submitted by} & \multicolumn{2}{c}{Number of pull requests integrated by} \\ 
         & Population        & Proportion(\%)       & Population               & Proportion(\%)              & Population                 & Proportion(\%)                \\ \cline{2-7} 
White    & 235541            & 47.8   \%              & 1543967                  & 61.57 \%                  & 1762204                    & 70.28   \%                      \\
API      & 33776             & 6.8   \%               & 162792                   & 6.49     \%                   & 122656                     & 4.89     \%                     \\
Hispanic & 12356             & 2.5    \%              & 64567                    & 2.57     \%                   & 74583                      & 2.97      \%                    \\
Black    & 638               & 0.1    \%              & 3092                     & 0.13      \%                  & 2090                       & 0.08      \%                    \\
AIAN     & 1                 & $\approx 0$    \%      & 3                        & $\approx 0$   \%              & 0                          & 0           \%                  \\
Unknown  & 210858            & 42.8      \%           & 733170                   & 29.24       \%                & 546058                     & 21.78      \%                   \\ \hline
Total    & 493170            & 100     \%             & 2507591                  & 100         \%                & 2507591                    & 100\% \\\hline \bottomrule                        
\end{tabular}
\end{table*}

From the $493,170$ developers in our dataset, while we classified the perceptible race and ethnicity of $282,312$ developers (57.2\%), $210,858$ (42.8\%) developers were classified as ``Unknown''. Furthermore, 96\% of the $250,759$ pull requests in our dataset were submitted by contributors outside the projects. 

Our demographic results are shown in Table~\ref{tab:eth_dist} where we can notice a low number of perceptible Non-White developers in GitHub (6.8\% perceptible API developers, 2.5\% perceptible Hispanic developers, 0.1\% perceptible Black developers, and $\approx 0\%$ perceptible AIAN developers) when compared to the number of perceptible White developers (47.8\%). Also, Table~\ref{tab:eth_dist} indicates that perceptible White developers submit six times more (61.57\%) pull requests than perceptible Non-White developers ($6.49 + 2.57 + 0.13 = 9.19$); and  perceptible White developers integrate ten times more (70.28\%) pull requests than perceptible Non-White developers ($4.89 + 2.97 + 0.08 = 7.94$). Finally, the results in Table~\ref{tab:eth_dist} indicate the percentage number of pull request contributions submitted and integrated by each perceptible race and ethnicity. While the vast majority of contributions (61.57\%) were submitted by developers perceptible as White, developers perceptible as API, Hispanic, Black, and AIAN, in total, have submitted less than 10\% of the contributions. 29.24\% of contributions were submitted by developers with ``Unknown'' perceptible race and ethnicity.

Table~\ref{fig:general_acc} contains the results after analyzing the general acceptance rate for each group. Table~\ref{fig:general_acc} indicates that while the highest acceptance rate among all perceptible races and ethnicities is that of perceptible White submitters with 82.6\%, the acceptance rate average of perceptible API, Hispanic, and Black submitters is 80.04\%, 81.59\%, and 81.34\%, respectively. Moreover, the lowest acceptance rate is that of ``Unknown'' perceptible race and ethnicity with 79.2\%. 

\begin{table}[h]
\centering
  \caption{Pull Request acceptance rate (represented in percent-age) per racial and ethnic group.}
  \label{fig:general_acc}
  \begin{tabular}{llllll}
    \toprule
    Status PR & Unknown & White & API & Hispanic & Black\\
    \midrule
    Rejected  & 20.8 & 17.4 & 19.96 & 18.41 & 18.66\\
    Accepted & 79.2 & 82.6 & 80.04 & 81.59 & 81.34\\ \hline
  \bottomrule
\end{tabular}
\end{table}

We also analyzed the distribution of developers and pull request acceptance rates between different groups when considering the integrator's perceptible race and ethnicity. For that, we calculated the acceptance rate of each developer against each integrator. We found that perceptible White submitters have an acceptance rate average of (73\%) when the integrator is also perceptible as White (but their acceptance rate average is (69\%) when measured against all integrators). The acceptance rate average of submitter-integrator pairs perceptible as API and Hispanic (when the submitter and integrator are perceptible to be in the same group) is 78\% and 76\%, respectively. In addition, we found that this average is higher than the cases where the integrator is perceptible as a White developer. This average is 69\% for perceptible API-White developer pairs and 71\% for perceptible Hispanic-White developer pairs. These differences are worthy of investigation because a preliminary analysis showed statistically significant differences between the pull request acceptance rate of each perceptible racial and ethnic group. We have added more details about these preliminary analyses in our online appendix (B)~\cite{repro}.

\begin{figure}[h]
\centering
  \includegraphics[width=\linewidth]{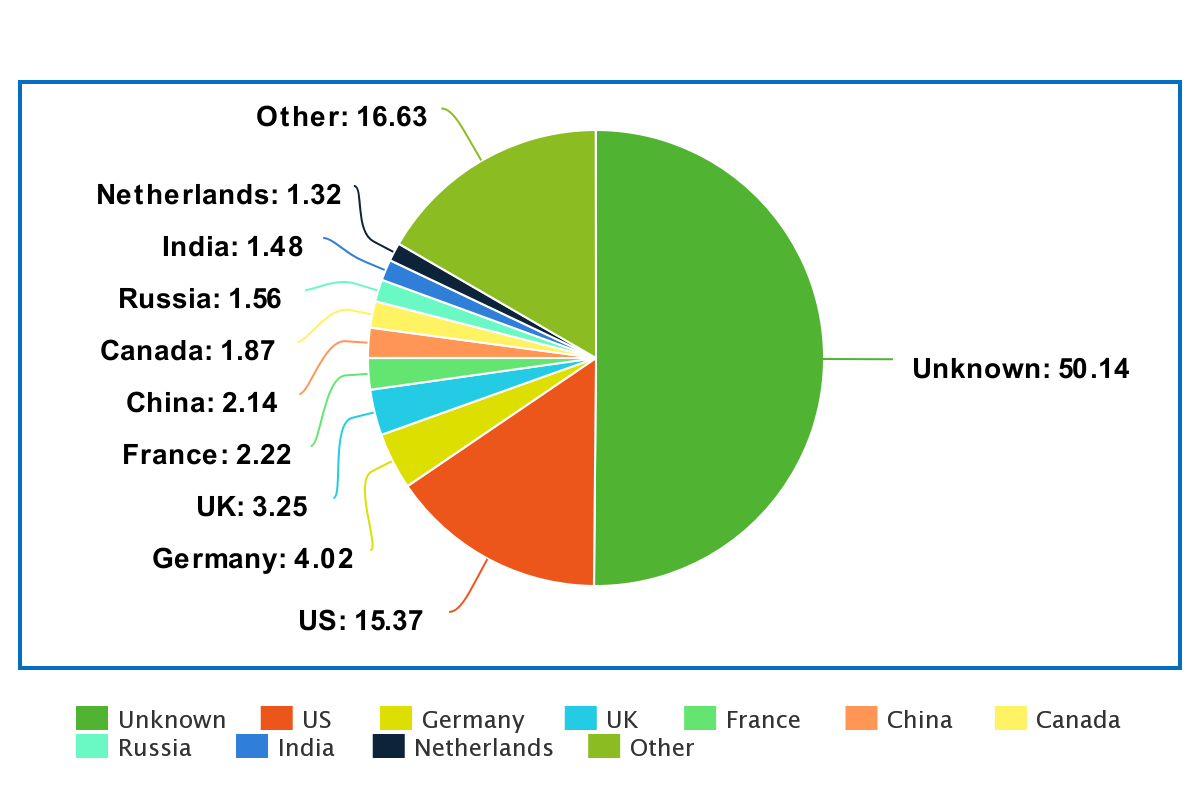}
\caption{Top Countries according to the number of developers}
\label{fig:top10} 
\end{figure}

Finally, we identified the country distribution of each perceptible racial and ethnic group in GitHub. Among $493,170$ developers, we identified the country of $245,881$ (49.85\%) of them. Figure~\ref{fig:top10} shows the top countries identified based on the number of developers. North American countries: U.S. and Canada, count for 17.24\% of developers. The top European countries: Germany, the UK, France, and the Netherlands, count for 13.03\% of developers. Russia, China, and India also appear in the top countries, with 5.18\% of developers. 

\begin{figure}[h]
  \includegraphics[width=\linewidth]{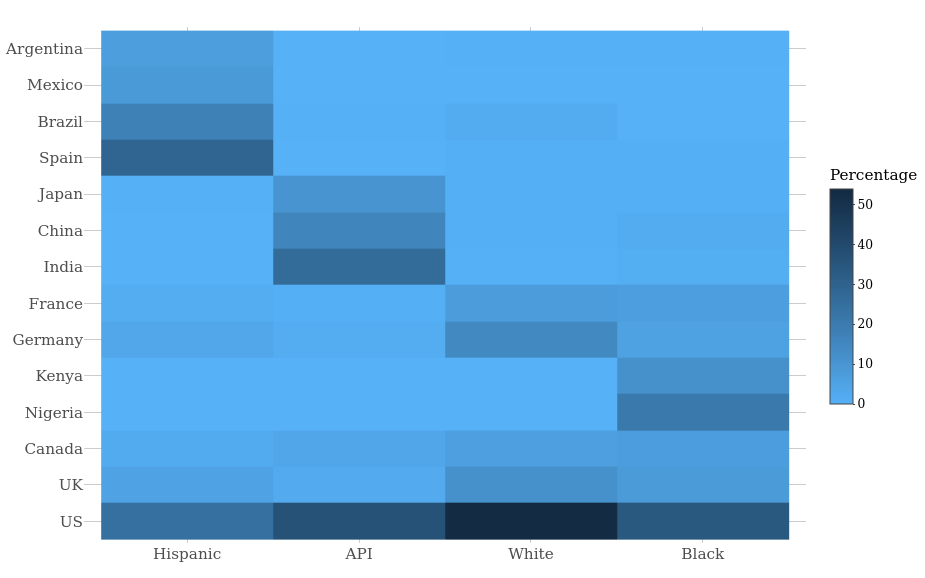}
\caption{perceptible race and ethnicity population proportion in the top countries}
\label{fig:heatmap} 
\end{figure}

We looked at the top five countries for each racial and ethnic group based on the number of developers as a soundness check for the perceptible race and ethnicity classification method. This process resulted in only fourteen distinct countries since three countries (U.S., UK, Canada) were among the top countries for different races and ethnicities. For the U.S., $87.09\%$ were perceptible as White users, $10.13\%$ were perceptible as API users, $2.59\%$ were perceptible as Hispanic users, and only $0.17\%$ were perceptible as Black users. For the UK, $93.56\%$ were perceptible as White users, $3.71\%$ were perceptible as API users, $2.52\%$ were perceptible as Hispanic users, and only $0.19\%$ were perceptible as Black users. In Canada, $88.23\%$ were perceptible as White users, $9.56\%$ were perceptible as API users, $1.88\%$ were perceptible as Hispanic users, and only $0.31\%$ were perceptible as Black users. Each cell in Figure~\ref{fig:heatmap} shows what proportion of the race and ethnicity resides in each country. The percentage numbers in Figure~\ref{fig:heatmap} are normalized using the IPFP (Iterative Proportional Fitting Procedure) function~\cite{fienberg1981iterative} of the TeachingSampling~\footnote{\url{https://rdrr.io/cran/TeachingSampling/man/IPFP.html}} package in R. From Figure~\ref{fig:heatmap}, we can observe that (1) the U.S. has the highest percentage of the perceptible Hispanic developers, perceptible API developers, perceptible White developers, and perceptible Black developers (the highest percentage in the U.S. is perceptible White developers); (2) besides the U.S., most of the perceptible API developers reside in Asian countries; (3) European and North American countries represent a higher proportion of perceptible White developers; (4) besides the U.S., South American countries and Spain represent a higher proportion of perceptible Hispanic developers; and (5) other than the U.S., African countries have a higher proportion of perceptible Black developers.

Figure \ref{fig:heatmap} shows that although the largest perceptible racial and ethnic group of developers in France and Germany is White, there are developers from these countries who are perceptible as Black developers and as Hispanic developers. Furthermore, for perceptible White countries such as Canada or U.S., the percentage of perceptible API developers is also considerable. 

\subsection{RQ1: To what extent is the submitter's perceptible race and ethnicity related to the acceptance probability of a pull request?}
\label{sec:rq2}



After excluding pull requests in which their submitter's perceptible race and ethnicity or integrator's perceptible race and ethnicity is unknown, we obtained $1,407,622$ pull requests from $187,616$ unique submitters (84.3\% perceptible White submitters, 11\% perceptible API submitters, 4.5\% perceptible Hispanic submitters, and 0.2\% perceptible Black submitters). We found no multicollinearity between all of the independent variables in all of our models as they have a value of less than 3. 

Table \ref{tab:tab_models} presents the coefficients of our independent and control variables in the mixed-effect logistic regression model, the $\rho\_value$, and the odds ratio. Each one of the coefficients in Table~\ref{tab:tab_models} reads as the effect of that variable on the log odds of pull request acceptance when other predictor variables are kept constant. The $\rho\_value$ shows the statistical significance of the variable to predict pull request acceptance. As in previous studies~\cite{rastogi2018relationship}, we use the odds ratio to understand the associations between our dependent and independent variables. The interpretation of the odds ratio for categorical and continuous predictors is different. While for categorical variables the odds ratio is compared to the default level of the categorical variable, for continuous variables the odds ratio indicates that the even is more likely to occur as the predictor increases or decreases. Notice that the default levels of our categorical variables, i.e., perceptible submitter race and ethnicity, submitter's continent, and same nationality, are perceptible White race and ethnicity, North America continent, and same nationality.

\begin{table}[h]
\renewcommand{\arraystretch}{1.3}
    
    \caption {Perceptible race and ethnicity related results of the mixed-effect models. $prs\_eth$ is the $submitter\_ethnicity$. Signif. codes: 0: ***; 0.001: **; 0.05:*}
     
    \label{tab:tab_models}
    \centering
    \begin{tabular}{l|l|l|l}
        \toprule
        - & Estimate & $\rho\_value$ & odd ratio \\
        \midrule
        (Intercept)               & 2.273e+00   & $<$ 2e-16 ***     & 9.70 \\
    repo\_maturity                & 8.872e-03   &  0.160673         & 1.01  \\
    repo\_popularity              & 0.000e+00   &  1.000000         & 1.00   \\
    repo\_team\_size              & 1.434e-02   &  0.148415         & 1.01  \\
    repo\_external\_contribs      & -2.608e-02  &  2.11e-07  ***    & 0.97  \\
    pr\_files\_changed            & -3.552e-04  &   $<$ 2e-16 ***   & 1.00   \\ 
    pr\_num\_comments             & 1.313e-03   &  0.648125         & 1.00  \\
    pr\_num\_commits              & -2.104e-01  &  $<$ 2e-16 ***    & 0.81  \\
    pr\_intra\_branch1            & 2.998e-01   &   $<$ 2e-16 ***   & 1.35 \\
    prs\_main\_member1            & 1.671e-01   &  1.92e-10  ***    & 1.18 \\
    prs\_popularity               & 9.346e-02   &   $<$ 2e-16 ***   & 1.10  \\
    prs\_watched\_repo1          & 8.225e-02    &  $<$ 2e-16 ***    & 1.09  \\
    prs\_tenure                   & 2.960e-02   &  4.82-10 ***      & 1.03 \\
    prs\_succ\_rate               & 1.820e-01   &   $<$ 2e-16 ***   & 1.20  \\ 
    prs\_experience             &-7.492e-02     &   $<$ 2e-16 ***   & 0.93  \\
    prs\_repo\_experience       & 5.102e-01 &   $<$ 2e-16 ***       &  1.67\\
    prs\_ethBlack               &-4.986e-02     &  0.604737         & 0.95  \\
    prs\_ethAPI                 &-1.042e-01     &  7.67e-12 ***     & 0.90  \\
    prs\_ethHispanic            &-6.522e-02     &  0.003408 **       & 0.94 \\
    prs\_continentAsia           &      -6.780e-02  &   0.000181 *** & 0.93  \\
    prs\_continentAfrica          &     -1.702e-01  &   0.005649 **  & 0.84  \\
    prs\_continentSouthAmerica   &     -5.146e-02  &   0.139946     & 0.95  \\
    prs\_continentAntarctica      &      2.056e-01  &   0.813285     & 1.23  \\
    prs\_continentUnknown         &      8.799e-02  &   9.75e-10 *** & 1.09   \\
    prs\_continentEurope          &      7.514e-02  &   1.34e-10 *** & 1.08  \\
    prs\_continentOceania         &      5.285e-02  &   0.056495    &  1.05 \\
    same\_nationalityDifferent.   & -2.105e-01  &   $<$ 2e-16 ***   &  0.81 \\
    same\_nationalityUnknown      & -2.045e-01  &   $<$  2e-16 ***  &  0.81 \\\hline 
    AIC                             & 965170.3   &    - & -                  \\             
    BIC                             & 965535.1   &    - & -                  \\ 
    Conditional R2                  & 0.07139062  &    - & -                  \\
    Marginal R2                     & 0.4768592  &    - & -                  \\  \hline
    \bottomrule
\end{tabular}
\end{table}

The results from Table~\ref{tab:tab_models} indicate that perceptible API submitters ($\rho\ < 0.0001 $) and perceptible Hispanic submitters ($\rho\ < .001 $) have statistically lower pull request acceptance rates than White submitters. Table~\ref{tab:tab_models} also shows that holding all the variables constant, the chances of a pull request being accepted when it is submitted by a perceptible API submitter and a perceptible Hispanic submitter have about 10\% ($exp(-0.1042)=0.90$) and 6\% ($exp(-0.06522)=0.94$) lower odds when compared to a pull request submitted by a perceptible White submitter, respectively. The number of perceptible Black submitters is very low in our dataset, therefore, we cannot safely interpret the odds ratio for perceptible Black submitters in GitHub as their $\rho\_value$ is not statistically significant. 

Other interesting results indicate that when control for all the other variables: (1) pull requests from submitters who reside in Asia and Africa have about 7\% and 16\% lower odds of being accepted when compared to submitters who reside in North America, respectively. However, pull requests from submitters who reside in Europe or who do not have the country information available have about 8\% and 9\% higher odds of being accepted when compared to submitters who reside in North America; (2) pull requests from submitters who are main team member have about 18\% higher odds of being accepted when compared with pull requests from developers who are not team members; (3) pull requests from submitters who have a different or unknown country of residence than the integrator have about 19\% lower odds of being accepted when compared with pull requests from submitters with the same country than the integrator; and (4) we expect to see about 20\% increase in the odds of a pull request being accepted for a one-unit increase in prs\_succ\_rate; similarly, we expect to see about 60\% increase in the odds of a pull request being accepted for a one-unit increase in the prs\_repo\_experience. However, for each one-unit increase in the prs\_experience, the likelihood that a pull request is being accepted by an integrator has about 6\% lower odds. 

We used the AIC~\cite{aic}, BIC~\cite{schwarz1978estimating}, conditional R2, and marginal R2 values to summarize the model presented in Table~\ref{tab:tab_models}. 
Then, we looked at the overall magnitude of the effects of all the variables in the model using the ANOVA statistical test \cite{cuevas2004anova}. As we expected, the effect size for the perceptible race and ethnicity is relatively small. We found that perceptible race ethnicity, can explain $0.6\%$ of data variance on pull request acceptance. This small effect size for the race and ethnicity related features is not strange as (1) the number of perceptible Non-White developers are small, and (2) there is a large proportion of pull request accepted and rejected by perceptible White and perceptible Unknown developers. Moreover, it is comparable to the effect size for the country information ($0.55\%$) in Rastogi~\emph{et al.}'s study~\cite{rastogi2018relationship}.

\subsection{RQ2:  To what extent is the submitter-integrator perceptible race and ethnicity pairs related to the acceptance probability of a pull request?}
\label{sec:rq4}

We built three mixed-effect regression models: (1) Same\_ethnicity\_Hispanic; (2) Same\_ethnicity\_API; and (3) Same\_ethnicity\_Black. In each model, we added a categorical variable $same\_ethnicity$ that indicates whether the integrator's estimated race and ethnicity is the same as the submitter's perceptible race and ethnicity. We did not remove any variable from the models as their $VIF$ were lower than 3.

\begin{table*}[htbp]
\centering
\renewcommand{\arraystretch}{1.2}
    \caption{Results of the the mixed-effect models when considering the integrator's perception of being of the same submitter's perceptible race and ethnicity. $same\_ethnicity$ is a categorical variable to measure if the estimated integrator's race and ethnicity and the perceptible submitter's race and ethnicity is the same. Note that $same\_ethnicity1$ indicated that the result of the variable is for the case when $same\_ethnicity = 1$ as $same\_ethnicity0$ is the default value. Signif. codes: 0: ***; 0.001: **; 0.05:*} 
    \label{tab:tab_models2}
    \centering
    \begin{tabular}{l|rrc|rrc|rrc}
    \toprule
    \multicolumn{1}{l|}{\multirow{2}{*}{-}} & \multicolumn{3}{c|}{Same\_ethnicity\_Hispanic} & \multicolumn{3}{c|}{Same\_ethnicity\_API} & \multicolumn{3}{c}{Same\_ethnicity\_Black} \\ \cline{2-10} \multicolumn{1}{c|}{} &
    \multicolumn{1}{c}{Coef. Estimate} & \multicolumn{1}{c}{ $\rho\_value$} & \multicolumn{1}{c|}{Odds} & \multicolumn{1}{c}{Coef. Estimate} & \multicolumn{1}{c}{$\rho\_value$} & \multicolumn{1}{c|}{Odds} & \multicolumn{1}{c}{Coef. Estimate} & \multicolumn{1}{c}{$\rho\_value$} & \multicolumn{1}{c}{Odds} \\
    \midrule
    (Intercept)      &  1.8770125  & $<$ 2e-16 ***& 6.533 & 1.833 & $<$ 2e-16 *** & 6.252 & 1.620779 & 7.56e-07 *** & 5.057\\
    repo\_maturity   &  0.0691582  & 0.010376 *  & 1.071 & 8.749e-02 & 6.66e-06 *** &1.091& 0.029331& 0.811438& 1.029\\
    repo\_popularity &  -0.0309107 & 0.424663     & 0.969 & -1.032e-01 & 0.000338 *** &0.901& -0.077749& 0.639436 &0.925\\
    repo\_team\_size &  -0.2244503 & 1.12e-07 *** & 0.798 & -2.038e-01 & 2.53e-10 *** &0.815&-0.376080 &0.024567 *&0.686 \\
    repo\_external\_contribs    & -0.0066519 & 0.768797 & 0.993 & 2.070e-02 & 0.216368 &1.020&0.189865& 0.056146 .&1.209\\
    prs\_succ\_rate             &  0.2155798 & $<$ 2e-16 *** & 1.240 &1.997e-01 & $<$ 2e-16 *** &1.221&0.218617& 0.036526 *&1.244\\
    pr\_files\_changed          & -0.0005260 & 0.000988 *** & 0.999 & -2.060e-04 & 4.01e-05 ***&0.999&0.001331& 0.544847&1.001\\
    prs\_main\_member1          & 0.5745823  & 0.001840 ** & 1.776 &7.273e-02 & 0.591753 &1.075&0.658779&0.521233& 1.932\\
    prs\_popularity             & 0.0031332  & 0.918167    & 1.003 &1.513e-01  & $<$ 2e-16 ***&1.163&0.321914&0.019400 *  &1.379 \\
    prs\_watched\_repo1         & 0.0542915  & 0.220059    & 1.055 & 1.026e-01 & 0.000265 ***&1.108&-0.206531& 0.362890  & 0.813\\
    prs\_tenure           & 0.1241617  & 1.91e-06 *** & 1.132 & 6.533e-02 & 2.22e-05 *** &1.067&-0.003738& 0.971614&0.996\\
    pr\_num\_comments     & 0.0122644  & 0.444758     & 1.0123 & 2.889e-02 & 0.003606 ** &1.029&0.106720& 0.185097 &1.112\\
    pr\_num\_commits      & -0.2260952 & $<$ 2e-16 ***& 0.797 & -2.184e-01 & $<$ 2e-16 ***&0.803&-0.297150& 4.21e-05 ***&0.742\\
    prs\_experience       & -0.2411955 & 2.14e-07 *** & 0.785 & -1.109e-01 & 0.000353 *** & 0.895&-0.222854&0.371192&0.800\\
    prs\_repo\_experience & 0.6165589  & $<$ 2e-16 ***& 1.852 & 4.867e-01 & $<$ 2e-16 *** &1.626& 0.891010& 0.000472 *** & 2.437\\
    prs\_continentAsia    & 0.4656812  & 0.034471 *  &1.593 & -1.032e-01 & 0.012374 * &0.901&-0.944570& 0.069703 .& 0.388\\
    prs\_continentAfrica  & 1.1283945  & 0.161496    &3.090 & -4.919e-01 & 0.261021 &0.611&0.118901 & 0.727461 & 1.126 \\
    prs\_contSouthAmerica  & -0.0315720 &  0.655180    & 0.968 & -3.510e-01 & 0.272887 &0.703 &0.698802 & 0.607235 & 2.011\\
    prs\_contAntartica     & - &  -    & - & 9.877e+00 & 0.975390  &19477& -&- &- \\
    prs\_contUnknown       & 0.3067880  &  0.000164 *** &1.359 & 6.934e-02 & 0.152619 &1.071 & 0.352854 & 0.349221 & 1.423\\
    prs\_contEurope        & 0.0802974  &  0.209706     &1.083 & -5.139e-02 & 0.482352 &0.949 &0.339520 & 0.322493& 1.404\\
    prs\_contOceania       & 0.1256501  &  0.673625     &1.133 & 8.654e-02 & 0.448014 &1.090 & -0.202291 & 0.793450& 0.816 \\
    same\_nationaDifferent  & -0.3137113 &  4.72e-06 *** &0.730 & 2.333e-01 & 3.37e-08 *** &1.262 &0.036415 & 0.913494 & 1.037 \\
    same\_nationaUnknown   & -0.4912372  &  6.39e-10 *** &0.611 & -2.516e-01 & 4.25e-08 *** &0.777 &0.303884 & 0.427581 &1.355\\
    intra\_branch1        & 0.4506343 & 1.40e-06 *** &1.569 & 6.014e-02 & 0.293974  &1.061 &0.361082 & 0.288162 & 1.434 \\
    \textbf{same\_ethnicity1}            & \textbf{0.5596182} & \textbf{1.19e-14 ***} &\textbf{1.750} & \textbf{3.065e-01} & \textbf{5.47e-15 ***} &\textbf{1.358} &\textbf{2.245506} & \textbf{0.047098 *} & \textbf{9.445} \\
        AIC               & 33862.3 &- & - & 84634.6 & - & - & 1442.3 & - & -      \\  
        BIC               & 34098.1 &- & - & 84905.0 & - & - & 1593.1 & - & -      \\ 
        Conditional R2    & 0.4536214 &- & - & 0.4850663 & - & - & 0.4478943 & - & -      \\
        Marginal R2       & 0.09629903 &- & - & 0.06846008 & - & - & 0.2087891 & - & -      \\  \hline
    \bottomrule
 \end{tabular}
 \end{table*}

Table \ref{tab:tab_models2} shows that when considering the new variable $same\_ethnicity$, the relationship between belonging to the same race and ethnicity group and pull request acceptance is statistically significant in all the three models, Same\_ethnicity\_Hispanic model, Same\_ethnicity\_API model, and Same\_ethnicity\_Black model.

Note that the reference value for $same\_ethnicity$ in the three models is 0. This means that the integrator and submitter race and ethnicity pair is not the same, e.g., the pull request was submitted by a perceptible Hispanic/API/Black submitter and it was closed by an estimated White integrator. To understand the association of $same\_ethnicity$ with the pull request acceptance when the estimated integrator and the perceptible submitter belong to the same race and ethnicity, we follow the standard practice of varying one while holding the other constant, and vice versa. Thus, when compared to estimated White integrators, the odds ratio is 1.75 for an estimated Hispanic integrator and perceptible Hispanic submitter, 1.36 for an estimated API integrator and perceptible API submitter, and 9.45 for an estimated Black integrator and perceptible Black submitter. This implies that (1) perceptible Hispanic submitters have 75\% higher odds of getting their pull requests accepted by estimated Hispanic integrators than by estimated White integrators; (2) perceptible API submitters have 36\% higher odds of getting their pull requests accepted by estimated API integrators than by estimated White integrators; and (3) perceptible Black submitters have 9 times\footnote{Note that in this case, we are not talking about \% change in odds.} higher odds of getting their pull requests accepted by estimated API integrators than by estimated White integrator.

The interesting result from Table \ref{tab:tab_models2} is that submitters with perceptible non-white races and ethnicities are more likely to get their pull requests accepted when the estimated integrator is from their same race and ethnicity rather than when the integrator is estimated to be White. Perceptible Hispanic submitters are more likely to get their pull request accepted when the integrator is estimated as Hispanic, perceptible API submitters are more likely to get their pull request accepted when the integrator is estimated as API, and perceptible Black submitters are more likely to get their pull request accepted when the integrator is estimated as Black.

We also looked at the overall magnitude of the effect of all the variables in the three models using the ANOVA statistical test \cite{cuevas2004anova}. We found that the $same\_ethnicity$ variable explains $6.749\%$ of data variance on pull request acceptance in Same\_ethnicity\_Hispanic,  $3.06\%$ of data variance on pull request acceptance in the Same\_ethnicity\_API model, and $4.5\%$ of data variance on pull request acceptance in the Same\_ethnicity\_Black model.

\section{Discussion}
\label{sec:disc}
\textbf{Less than 10\% of contributions that are finally integrated into OSS projects are from perceptible Non-White developers.}

Among $493,170$ developers, we identified that almost half (47.8\%) were perceptible as White developers. This result is consistent with Rastogi~\emph{et al.}'s work~\cite{rastogi2018relationship} who found that among the top seventeen countries contributing to OSS in GitHub, twelve are in Europe or North America and they contribute with 72\% of pull requests. Therefore, OSS contributions in GitHub might be dominantly from perceptible White developers. Contrary to Rastogi~\emph{et al.}'s work~\cite{rastogi2018relationship}, we decided to use the continent of residence of the developers to mitigate the imbalanced data issues relating to using country of residence. Thus, we might have missed differences in pull request acceptance rates between perceived races and ethnicities and different countries in the same continent. We believe that there is a lot of room to understand the reasons for these results. 

Furthermore, our results show a low number of perceptible Non-White developers in OSS when compared to perceptible White developers. In GitHub, only 6.8\% of OSS contributors were perceptible as API developers, 2.5\% of OSS contributors were perceptible as Hispanic developers, and 0.1\% of OSS contributors were perceptible as Black developers. From our results, we know that 0.13\% of the contributions submitted to OSS projects in GitHub were submitted by perceptible Black developers. We believe that this low percentage may indicate the omission of a large proportion of the human population. 

When filtering these results per country, we found that perceptible API submitters, Hispanic submitters, and Black submitters in the U.S. account for 9.3\%, 2.5\%, and 0.2\%. According to \emph{The Bureau of Labor Statistics}\footnote{\url{https://www.bls.gov/cps/cpsaat11.html}}, API programmers, Hispanic programmers, and Black programmers account for 37.7\%, 5.1\%, and 5.8\% of the programmers working in the U.S. respectively.  
These results are significantly higher than our findings and indicate that while the majority of OSS developers in the U.S are perceptible as White, perceptible API, Hispanic, and Black communities might not be proportionally represented in GitHub. 

We also found that contributions from submitters in non-White continents such as Asia and Africa have about 7\% and 16\% lower odds of being accepted when compared to contributions from submitters in North America. However, contributions from submitters in Europe, a historically White continent, have about 8\% higher odds of being accepted when compared to submitters who reside in North America.

These previous results indicate that GitHub should continue and even expand their efforts \footnote{\url{https://github.com/about/diversity/report}} to get more perceptible Black, AIAN, and 2PRACE developers engaged in the platform. Also, the OSS community needs to do more to include more perceptible Black, AIAN, and 2PRACE voices in their projects. The 2019 annual Octoverse report~\cite{octoverse} stated that the development of source code in 2019 was more global as the number of non-white communities grow across Asia and Africa. Therefore, the current trend may change and GitHub would have more contributions from perceptible Black submitters. If this happens, OSS communities should work to foster a healthier environment where all submitters can feel that their contributions are evaluated fairly. 

We believe that the underrepresentation of large proportions of the human population (i.e, Hispanic developers and Black developers) may have unwanted consequences for OSS communities. For example, it may cause not only lack of diversity and critical opinions from a very large population of the world, but also may prevent Non-White developers from contributing to OSS which would make the OSS development by and for White developers.

In contrast to the small percentages for perceptible Non-Whites developers when compared to perceptible White developers in OSS in GitHub, we found that a large proportion of the developers' race and ethnicity (42.8\%) was perceptible as Unknown. We hypothesize that this high percentage may be caused for (1) users that may prefer using an alias rather than sharing their real name; and (2) errors in the estimation of some races and ethnicities when using the Name-prism tool. 

We also found that submitters perceptible as Hispanic an API have lower odds (6-10\%) of getting their pull requests accepted when compared to perceptible White submitters. Hence, we think that the SE research community should investigate how to foster a more diverse OSS community and identify potential barriers that prevent Non-White developers from participating in OSS.

Besides the effort that researchers should do, practitioners also should be aware that some developers might find barriers when contributing to OSS in GitHub and speak up if they witness bad behavior and if they feel safe. The Open Source Survey in 2017 reported that around 50\% of GitHub's respondents had witnessed bad behavior in Open Source. This survey found that about 11\% of total respondents and 3\% of experienced respondents have witnessed stereotyping as a negative behavior.

If such stereotyping behavior exists against API developers, Hispanic developers, and Black developers, they might stop contributing to OSS communities. The authors believe that if this happens, it might affect the role of these developers in the tech community as a whole. According to the Open Source Survey in 2017~\cite{survey}, half of the respondents stated that their OSS contributions were a crucial factor for launching their professional careers. Therefore, OSS communities and GitHub should avoid any possible discrimination against developers' perceptible race and ethnicity.\\

\textbf{Being perceptible as a White submitter has a positive relation on pull request acceptance probability, but being perceptible as an API or Hispanic submitter does not}.

We found a statistically significant relationship between pull request acceptance and submitter's perceptible race and ethnicity. These results are in accordance with previous studies, which found that non-technical factors such as gender~\cite{vasilescu2015gender}, personality traits~\cite{iyer2019effects}, or the country of residence~\cite{rastogi2018relationship} are correlated to pull request acceptance probability.

Table~\ref{tab:eth_dist} shows that perceptible White developers submit six times more pull requests than perceptible Non-White developers, and that perceptible White developers integrate ten times more pull requests than perceptible Non-White developers. While these two findings already indicate a serious diversity issue with OSS communities, the results from Table~\ref{tab:eth_dist} also indicate that when it comes to positions of power (integrator) in an OSS environment there is an even higher lack of diversity. Notice that the percentage of estimated White integrators is higher (by 8.7\%) than the percentage of perceptible White submitters. This percentage is lower (by 1.6\%) for perceptible API submitters, and it is almost the same for perceptible Black submitters and perceptible Hispanic submitters. Therefore, we believe that perceptible Black, Hispanic, and API developers are not proportionally represented in positions of power in OSS. 

It is important to mention that the results from Table~\ref{tab:tab_models} and Table~\ref{tab:tab_models2} may contradict the intention of OSS communities to behave as a meritocracy because OSS integrators in GitHub may consider submitters' non-technical features (i.e, perceptible race and ethnicity, continent of residence, or tenure) as important factors to accept pull requests instead of just looking at the quality of the contribution. For example, integrators may promote a way to unconsciously/unintentionally bias towards more experienced developers. We have not examined the relationship between experience and race and ethnicity. This is a possible future exploration. Our study, therefore, is providing initial insights into possible unconscious or unintentional racial and ethnic biases that may exist in OSS development.

We believe that a single or double-blind pull request review process in GitHub might help to mitigate potential unconscious biases. In the academic world, it is common to use a single-blind review process in journals or a double-blind review process in conferences as there is consistent evidence on bias related to the authors' institution prestige~\cite{pontille2014blind}. Furthermore, it has been shown the preferential treatment reviewers have for authors from English speaking countries~\cite{ross2006effect} and authors from the same country~\cite{daniel1993guardians,ernst1991chauvinism}.

\section{Threats to Validity}
\label{sec:thre}
We present our validity threats in terms of the four main threats in empirical software engineering research~\cite{wohlin2012experimentation}.

\subsection{Construct Validity}
We want to study the population of software developers in GitHub. Thus, to remove bias from projects that were not related to software development activities, we have used the reaper tool~\cite{munaiah2017curating} to classify engineered and non-engineered projects. This classification might introduce limitations to our results if there are some software developers who work on projects that we classified as non-engineered projects.

Some previous studies~\cite{rastogi2018relationship,zhou2019fork,gousios2014exploratory} identified pull request status using a set of heuristics, but we extracted this data using GitHub's API directly. Our approach, present two limitations: (1) there is no difference between pull requests that are cherry-picked\footnote{we refer to cherry-picked when not all commits in a pull request are accepted and merged. There might be some modifications by the integrator before merging the pull request} and the ones that are rejected; and (2) some merged pull requests might be classified as non merged because they did not leave traces in GitHub's API. 

While we cannot do anything to mitigate the first limitation, we can mitigate the second by manually analyzed 557 random pull requests that were classified as non-merged. In this analysis, we identified whether the pull requests ended up being merged in the project. Then, from the pull requests that ended up being merged, we identified if the heuristics proposed by Gousios et al.~\cite{gousios2014exploratory} and Zhou et al.~\cite{zhou2019fork} are able to automatically identify them. Finally, if the heuristics can identify the pull requests as merged, we analyzed whether the submitter got credit for the contribution by looking at the authorship of the source code that was merged in the project. First, two of the authors classified the 557 pull request as ``true rejected'' or ``finally merged'' looking at the comments and the workflow reflected in each pull request. The results indicate that 406 pull requests (73\%) were ``true rejected'' and 152 (27\%) ended up as merged in the projects. The second author manually applied the heuristics proposed by Gousios et al.~\cite{gousios2014exploratory} and Zhou et al.~\cite{zhou2019fork} to these 152 pull requests and found that the heuristics can identify 105 out of 152 pull requests. Finally, the second author identified if the submitter got credit for the contribution. From the 105 pull requests, 47 (8\%) pull requests gave credit to the submitters, 25 (4\%) pull requests did not give credit to the submitters, and 33 (6\%) pull requests were unsure cases as there was no commit attached to the pull request. 

Since our study's goal was to not only identify the contributions that were merged but also whether the submitters got credit for them, we believe that including such heuristics~\cite{zhou2019fork,gousios2014exploratory} could be worse for the validity of the conclusions than useful as these heuristics would include 10\% (58 out of 557) of pull requests where their submitter did not get clear credit.


\subsection{Internal Validity} 
Identifying perceptible races and ethnicities using names is ongoing research, which can be further explored. Name-Prism~\cite{ye2017nationality} may identify misassigned races and ethnicities, but it has been evaluated in previous studies in~\cite{alshebli2018preeminence}, and it presents an F1 score of 0.795. Furthermore, Name-Prism~\cite{ye2017nationality} uses US-based racial and ethnic categorization which may be a threat because predominant races and ethnicities may vary depending on the country. However, this tool is trained on a 74M labeled name set from 118 countries around the world, therefore, this categorization represents the biggest races and ethnicities in the world.

Providing a name is not mandatory on GitHub. Thus, users can omit their names or use other than their our, which may affect our results. However, our paper studies whether there is a relationship between the developer's perceptible race and ethnicity and the evaluation process of the pull requests. 

We only analyze what OSS developers could perceive as the race and ethnicity of another developer from their name, in the absence of any indicator on GitHub. This paper does not analyze whether other's perception of one's race and ethnicity is more important than their actual race and ethnicity. Therefore, in our study, it is not essential whether the developers are not using their real names. If they use a name associated with a different perceptible race and ethnicity other than their own, then any other developer would perceive the race and ethnicity derived from their chosen name, much like the tool we use. In addition, we used Stanford NERTagger to distinguish between human names and other groups of words. 

Another internal threat might be the class imbalance of our independent race and ethnicity variable. 



\subsection{External Validity}
Even though our dataset is bigger than previous studies~\cite{tsay2014influence,rastogi2018relationship,gousios2014dataset}, it is not representative of the whole community. Many GitHub users have unknown accounts, which makes it difficult to draw any conclusion. However, considering that other's perception of one's race and ethnicity is essential in our study, users with unknown accounts help us to investigate whether perceptible race and ethnicity from GitHub names affects acceptance probability.

\subsection{Conclusion Validity}
Although we captured most of the independent variables in the literature, there may be other independent variables that we have missed. 

Our study analyses whether the integrator identifies as being of the same perceptible race and ethnicity of the submitters. While it would be ideal to know the racial and ethnic identity of the integrators, we just are able to know their race and ethnicity as estimated by NamePrism. Although this is a threat, we would like to restate that in our evaluation of NamePrism, the accuracy of identifying race from their name using this technique was high. Also, the F1 score of the NamePrism tool is high (0.795).

We believe that research should actively look for other features that might affect pull request acceptance. Our findings show a correlation between perceptible race and ethnicity and pull request acceptance. However, we cannot claim that this is due to the existence of any racial and ethnic discrimination. Our approach uses only six racial and ethnic groups, which might not be a good representative of all races and ethnicities. However, these groups are considered to include the majority of the population of the world.

\section{Conclusion and Future Work}
\label{sec:conc}
Usually, OSS projects are the result of many collaborations from diverse developers with different backgrounds. Although the acceptance of such contributions should be based on the quality of the source code being contributed~\cite{scacchi2007free}, recent studies have shown that diversity issues have significant positive and negative effects on the acceptance or rejection of these contributions~\cite{tsay2014influence,iyer2019effects,vasilescu2015gender,terrell2017gender,rastogi2018relationship}. Therefore, this paper assists with the first empirical study that analyzes how perceptible race and ethnicity relates to the evaluation of the OSS contributions in GitHub.

We analyzed more than four million pull requests from $493,170$ OSS developers in GitHub. We first identified developers' perceptible race and ethnicity based on their GitHub names using the Name-Prism tool~\cite{ye2017nationality}. We then linked the developers' perceptible race and ethnicity with their pull requests, and we finally built regression models to study the relationship between developers' perceptible race and ethnicity, and pull request acceptance probability.

Our findings indicate a low number of contributions that are currently integrated into OSS projects from perceptible Black developers (0.1\%), Hispanic developers (2.5\%), and API developers (6.8\%). We believe that these low numbers of perceptible Non-White OSS developers in GitHub can be seen as a call for action, because without diversity in OSS development some racial and ethnic groups may stop contributing to OSS projects. It could cause not only a larger under-representation of OSS developers in GitHub but also that OSS products are created by and for White people.

Furthermore, our findings also indicate that the relationship of perceptible API and perceptible Hispanic developers on pull request acceptance is statistically lower compared to the relationship of perceptible White developers. Overall, we find differences between the evaluation of perceptible White contributors and perceptible Non-White contributions. Therefore, these results may indicate that OSS communities need to be aware of the diversity of their developers and move towards a more racially and ethnically diverse OSS community. 

Although our quantitative results are a first step to be aware of the perceptible race and ethnicity problem in OSS, further research should be done. For example, researchers should investigate how to foster a more diverse OSS community and identify whether there are barriers that prevent Non-White OSS developers from participating in GitHub. Also, a thorough qualitative survey to support our quantitative results can be done. Another line of research is developing new tools that help OSS communities to behave as true meritocracies and tools that allow developers to speak out against stereotyping behavior when they see it. For example, integrated tools in GitHub that allow a single-blind code review process for submitters. That way, integrators can focus just on the quality of the source code contributed by the submitter. 
These tools can help OSS projects by fostering a healthier OSS community. 
 
\subsection{Replication package}
Supplementary material associated with this article as well as the replication package can be found in https://github.com/uw-swag/2020-TSE-Developers-Perceptible-Ethnicity-and-PR-evaluation.

\ifCLASSOPTIONcompsoc
  \section*{Acknowledgments}
\else
  \section*{Acknowledgment}
\fi
We thank NSERC and SSHRC for their funding that supported this research. We also would like to thank, Dr. Audris Mockus, Dr. Shane McIntosh, and everyone in the SWAG lab for their continuous feedback. Finally, we would like to thank the Associate Editor (Dr. Martin Robillard) and anonymous reviewers for helping us improve this manuscript.

\bibliographystyle{IEEEtran}
\bibliography{references}

\end{document}